\documentclass[aps,prd,showpacs,preprint,onecolumn,floatfix,nofootinbib]{revtex4}

\newcommand{\be}{\begin{equation}}
\newcommand{\ee}{\end{equation}}
\newcommand{\ba}{\begin{eqnarray}}
\newcommand{\ea}{\end{eqnarray}}
\newcommand{\bc}{\begin{center}}
\newcommand{\ec}{\end{center}}

\usepackage{amsmath}
\usepackage{amstext}
\usepackage{amsthm}
\usepackage{amssymb}

\begin{document}

\preprint{MZ-TH/04-16}

\begin{center}

{\LARGE \textsc{Proper Time Flow Equation for Gravity}}

\vspace{1.4cm}
{\large A.~Bonanno}\\

\vspace{0.7cm}
\noindent
\textit{INAF -  Osservatorio Astrofisico di Catania,
Via S.Sofia 78, I-95123 Catania, Italy\\
INFN, Via S. Sofia 64, I-95123 Catania, Italy}\\

\vspace{0.7cm}
{\large M.~Reuter}\\

\vspace{0.7cm}
\noindent
\textit{Institute of Physics, University of Mainz\\
Staudingerweg 7, D-55099 Mainz, Germany}\\

\end{center}

\vspace*{0.6cm}

\date{\today}

\begin{abstract}
We analyze a proper time renormalization group equation for Quantum Einstein Gravity
in the Einstein-Hilbert truncation and compare its predictions to those of the
conceptually different exact renormalization group equation of the effective average action. We employ a smooth infrared
regulator of a special type which is known to give rise to extremely precise critical
exponents in scalar theories. We find perfect consistency between the proper time and the average action
renormalization group equations. In particular the proper time equation, too, predicts the existence of a non-Gaussian fixed point
as it is necessary for the conjectured nonperturbative renormalizability of Quantum Einstein Gravity.
\end{abstract}

\maketitle

\section{\label{sec:level1}Introduction}
Recently a lot of work went into the exploration of the Wilsonian renormalization group 
(RG) flow of Quantum Einstein Gravity 
\cite{mr,dou,oliver1,frank1,oliver2,souma,percaper,frank2,litimgrav} 
and its possible impact on black holes and 
cosmology 
\cite{bh,cosmo1,cosmo2,elo,esposito,h1,h2}.
By definition, Quantum Einstein Gravity (QEG) is the conjectured quantum field
theory of the spacetime metric whose bare action is (infinitesimally close to) an ultraviolet (UV) 
attractive non-Gaussian fixed point of the RG flow. As outlined by S.Weinberg  long ago \cite{wein}
this theory would be nonperturbatively renormalizable or ``asymptotically safe" and, most probably, 
predictive and internally consistent even at the shortest sub-Planckian distances. Most of the 
existing RG calculations  were performed within the framework of the exact RG equation pertaining to the 
effective average action \cite{avact,avactrev,ym}, 
applied to a truncated space of action functionals (theory space).
It turned out that the UV fixed point necessary for asymptotic safety does indeed exist within the 
Einstein-Hilbert \cite{souma,oliver1,frank1} and the $R^2$-truncation \cite{oliver2}, and 
detailed analyses of the reliability of those approximations \cite{oliver1,oliver2} revealed that the 
fixed point found is unlikely to be a truncation artifact. But clearly it is necessary 
to collect further evidence for its existence at the exact level, both within the effective 
average action formalism (by generalizing the truncation) and with conceptually  independent 
methods. 

In fact, a conceptually independent support of the asymptotic safety conjecture
comes form the work of Forg\'{a}cs and Niedermaier \cite{max} who were able to analyze the path integral 
over the metrics with two Killing vectors exactly and to show that at least this sector is asymptotically 
safe. In view of recent progress \cite{ajl} on causal, {\it i.e.} Lorentzian dynamical triangulations
\cite{amb} one also might hope that it will soon be possible to compare the analytical results
to Monte Carlo data. 

In the present paper we are going to analyze the non-Gaussian fixed point of QEG using the ``proper time
renormalization group" which has been extensively used for other theories recently 
\cite{pt,wf1,litim,sm}.
While this formalism is less complete and systematic than the exact RG equation of the average action
\cite{ym}, it has had quite spectacular successes in high precision calculations of critical exponents
\cite{wf1}. Even though, at the technical level, the proper time method is reminiscent of the 
average action approach, it is well known \cite{litim} that the proper time RG equation is {\it not} a special
case of the exact RG equation of the average action; it is conceptually independent in this sense. 
The approach we are going to use implements a certain RG improvement of one-loop 
perturbation-theory, employing an infrared (IR)-regularized
proper time representation of the corresponding one-loop determinants. We shall derive a flow equation
for the scale dependent gravitational action from it whose structure is quite different from that of the 
exact RG equation of the average action, for every possible choice of its built-in cutoff operator 
${\cal R}_k$ \cite{avactrev}. Apart from the conceptual independence the main motivation for the present
analysis are the remarkably precise critical exponents which have been obtained for the Wilson-Fisher
fixed point with this method \cite{wf1}. In fact, we shall be mostly interested in the universal properties of the 
non-Gaussian fixed point of QEG,  $ (g_\ast, \lambda_\ast)$, in particular its critical exponents $\theta'$ and $\theta''$ and the 
product $g_\ast \lambda_\ast$.  

\section{The proper time flow equation}
The ultimate goal of the RG approach would be the calculation of the integral 
$Z=\int {\cal D}\gamma_{\mu\nu} \exp(-S[\gamma_{\mu\nu}])$ over all metrics
$\gamma_{\mu\nu}$ where $S$ is  an arbitrary bare action, invariant under general coordinate
transformations. We are going to employ the background 
field formalism and start by writing $\gamma_{\mu\nu}$ as a background 
$\bar{g}_{\mu\nu}$ plus a (not necessarily small)
fluctuation: $\gamma_{\mu\nu}=\bar{g}_{\mu\nu}+h_{\mu\nu}$. Furthermore, we gauge fix the path integral 
by adding gauge fixing and ghost terms to $S$. We use an a priori arbitrary gauge fixing term 
$S_{\rm gf}[h;g]$ of the background-type, meaning that it is invariant under a {\it combined}
transformation of $h_{\mu\nu}$ and $\bar{g}_{\mu\nu}$. This property will guarantee that the 
effective action becomes invariant under general coordinates transformation. Thus, denoting the Faddeev-Popov
ghosts by $C^{\mu}$ and $\bar{C}_\mu$, and their action by $S_{\rm gh}[h,C,\bar{C};\bar{g}]$, the path
integral reads $Z=\int{\cal D}h{\cal D}C{\cal D}\bar{C} \exp ( -S[\bar{g}+h]-S_{\rm gf}-S_{\rm gh})$. Next
we couple $h_{\mu\nu}$ to a source, Legendre transform $\ln Z$ in the usual way, and thus obtain the 
effective action $\Gamma[\bar{h};\bar{g}]$. Here $\bar{h}_{\mu\nu}\equiv \langle h_{\mu\nu} \rangle$ is the expectation
value of the fluctuation. The corresponding expectation value of the complete metric will be
obtained by $g_{\mu\nu}= \langle \gamma_{\mu\nu} \rangle \equiv \bar{g}_{\mu\nu} +\bar{h}_{\mu\nu}$. It is
customary to consider $\Gamma$ a functional of $g$ and $\bar{g}$ rather than $\bar{h}$ and $\bar{g}$,
and to define $\Gamma[g,\bar{g}]\equiv \Gamma[\bar{h}=g-\bar{g}; \bar{g}]$. As always in the background
formalism \cite{mr,adler,ym} the ``ordinary" effective action, the generating functional of one-particle
irreducible Green's functions, $\Gamma[g]$, 
is obtained by setting $\bar{h}=0$ or $\bar{g}=g$: $\Gamma[g]=\Gamma[g,g]\equiv\Gamma[\bar{h}=0;\bar{g}]$.

As in every field theory, the one-loop correction to the effective action is given by a set 
of determinants: 
\be\label{1}
\Gamma_1[g,\bar{g}] = \frac{1}{2}{\ln \det} \; \widehat{S}^{(2)} -{\ln \det} \; S_{\rm gh}^{(2)}
\ee
Here we set $\widehat{S}[\bar{h};g]\equiv S[\bar{g}+\bar{h}] +S_{\rm gf}[\bar{h}; \bar{g]}$
and $\widehat{S}^{(2)}$ denotes the matrix of the second functional derivatives of 
$\widehat{S}$ with respect to $\bar{h}_{\mu\nu}$, and likewise for the ghosts.

In order to evaluate $\ln \det \Omega ={\rm Tr}\ln \Omega$ for $\Omega = \widehat{S}^{(2)}$ and 
$S_{\rm gh}^{(2)}$ we use the proper time representation ${\rm Tr} \ln \Omega = -\int_0^\infty (ds/s) \; {\rm Tr }\exp (-s\Omega)$.
We regularize the short-distance singularities of this expression by introducing a UV cutoff $\Lambda$ 
which suppresses the contributions to the $s$-integral with $s \lesssim \Lambda^{-2}$. Furthermore, in 
order to set up a Wilsonian RG equation, we also introduce a IR cutoff $k$ which suppresses the contributions due to 
large proper  times $s\gtrsim k^{-2}$. In principle one could employ a sharp proper time cutoff which amounts
to replacing $\int_0^\infty ds$ with $\int_{1/\Lambda^2}^{1/k^2} ds$. 
In this paper we use instead a more general class of 
proper time cutoffs.
One writes 
${\rm Tr} [\ln \Omega]_{\rm reg} = -{\rm Tr} \int_0^\infty f_k(s) \exp (-s\Omega)$ where the shape of the function
$f_k(s)$ is chosen such that it interpolates smoothly between $f_k(s)\approx 0$ for $s\gg k^{-2}$ and $f_k(s)\approx 1$
for $s\ll k^{-2}$, and similarly in the UV. Actually the UV cutoff will play no role in that follows since we shall 
concentrate on the $k$-derivative of the regularized determinant. Introducing the ``RG-time" $t=\ln k$ we have
\be\label{2}
\partial_t {\rm Tr}[\ln \Omega]_{\rm reg} = -{\rm Tr}\int_0^\infty\frac{ds}{s}\; \partial_t f_k(s) \exp (-s\Omega)
\ee
For an appropriately chosen $f_k(s)$ the scale derivative $\partial_t f_k(s)$ is independent of $\Lambda$.
Hence, at the level of (\ref{2}), we may safely perform the limit $\Lambda \rightarrow \infty$. 

For actual calculations  we shall use the following one-parameter family of smooth 
cutoffs $f_k\equiv f_k^{m}$ which has been extensively used in the literature \cite{pt},
in particular in high precision calculation 
of the critical exponents\footnote{We do not 
perform the rescaling $k^2\rightarrow mk^2$ as in \cite{sm}.} \cite{wf1}:
\be\label{3}
f_k^m(s) = \frac{\Gamma(m+1,{\cal Z} sk^2) -\Gamma(m+1,{\cal Z} s\Lambda^2)}{\Gamma(m+1)}
\ee 
Here $m$ is an arbitrary real, positive parameter which controls the shape of the $f_k^m$ in the 
interpolating regions, and $\Gamma(\alpha,x)=\int^\infty_x dt \; t^{\alpha-1} e^{-t}$ denotes the incomplete 
Gamma-function. Furthermore, ${\cal Z}$ is a constant (actually a matrix in field space) which 
has to be adjusted in the following way.      
If the field of type ``$a$" has a kinetic term $-Z_a D^2$, we set ${\cal Z} =Z_a$ for this field 
to make sure that the eigenvalues of $-D^2$ are cut off precisely at $k^2$, rather than $k^2/{\cal Z}_a$ \cite{avactrev}.
The scale derivative of (\ref{3}) is given by 
\be\label{4}
\partial_t f_k^m(s) = -2\;\frac{1}{m!}\; ({\cal Z} s k^2)^{m+1} \exp (-{\cal Z} sk^2) 
\ee
which is indeed independent of $\Lambda$. Inserting (\ref{4}) into (\ref{2}) and performing the integral
we obtain
\be\label{5}
\partial_t {\rm Tr} [{\rm \ln} \Omega]_{\rm reg} = 
2 \; {\rm Tr} \Big ( \frac{{\cal Z}k^2}{\Omega +{\cal Z}k^2} \Big )^{m+1} 
\ee
For $m$ integer, in the limit $\Lambda \rightarrow \infty$, Eq.(\ref{3}) yields the following explicit 
representation:
\be\label{6}
f_k^m(s) = \exp (-{\cal Z} sk^2 ) \sum_{\mu =0}^{m} \frac{1}{\mu!} \; ({\cal Z} sk^2)^{\mu}
\ee

If we regularize the determinants in (\ref{1}) with a proper time cutoff then $\Gamma_1$ and, as a consequence, 
the complete one loop effective action becomes $k$-dependent. We denote it by $\widehat{S}_k[g,\bar{g}]\equiv
\widehat{S}[\bar{h};\bar{g}] + \Gamma_1[g,\bar{g}]_{\rm reg}$. Its scale derivative is given by
\be\label{7}
\partial_t \widehat{S}_k[g,\bar{g}] = -\frac{1}{2} {\rm Tr} \int_0^\infty \frac{ds}{s}
\; \partial_t f_k^m(s) [\exp(-s\widehat{S}^{(2)}) -2\exp(-s{S}^{(2)}_{\rm gh}) ]
\ee 

Up to this point we performed a standard one-loop calculation, with the classical action $\widehat{S}$
appearing on the RHS of (\ref{7}), and the quantum corrected $\widehat{S}_k$ on its LHS. 
Now we perform a RG-improvement of this one-loop calculation by feeding-back $\widehat{S}_k$ into the RHS
of (\ref{7}). This leads to the following ``proper time RG equation" for the functional $\widehat{S}_k$:
\be\label{8}
\partial_t \widehat{S}_k [g,\bar{g}] = -\frac{1}{2} {\rm Tr} \int_0^{\infty} \frac{ds}{s}\; 
\partial_t f_k^m (s) [\exp(-s\widehat{S}^{(2)}_k)-2\exp (-s{S}^{(2)}_{\rm gh} ) ]
\ee
For a detailed discussion of flow equations of this type we refer to the literature \cite{pt,wf1}.
Note that only {\it after} the derivatives implicit in $\widehat{S}_k^{(2)}$
have been performed one can set $g=\bar{g}$ in order to obtain the ``ordinary" running action 
$S_k[g]\equiv \widehat{S}_k[g,g]$ since these derivatives are to be performed at fixed 
$\bar{g}$. 

The above flow equation could be generalized to allow for an RG evolution in the ghost
sector. Starting from the standard one-loop effective action 
$\Gamma_1[g,\bar{g},\xi,\bar{\xi}]_{\rm reg}$
in presence of nonvanishing ghost expectation values 
$\xi=\langle C \rangle$ and $\bar{\xi} = \langle \bar{C} \rangle$,
and following the same steps as above, we arrive at
\be\label{+}
\partial_t \widetilde{S}_k [g,\bar{g},\xi,\bar{\xi}]
=-\frac{1}{2} \; {\rm STr} \int_0^{\infty} \frac{ds}{s}\;
\partial_t f_k^m(s)  \exp(-s\widetilde{S}^{(2)}_k) 
\ee
The functional
$\widetilde{S}_k [g,\bar{g},\xi,\bar{\xi}]$ is a generalization of $\widehat{S}+{S}_{\rm gh}$ at the tree-, and
$\widehat{S}+{S}_{\rm gh}+\Gamma_1[g,\bar{g},\xi,\bar{\xi}]_{\rm reg}$ at the one loop level,
respectively. In (\ref{+}), $\widetilde{S}^{(2)}_k$ is a supermatrix involving derivatives with respect to
$g$, $\xi$ and $\bar{\xi}$, and the supertrace takes care of the relative minus sign between the graviton and
the ghost contributions. In the present paper we are not going to use the generalized flow equation (\ref{+}).

\section{The $\beta$-functions of $g$ and $\lambda$} 
The RG equation (\ref{8}) describes a flow on the infinite dimensional space of functionals $\widehat{S}[g,\bar{g}]$.
We can try to obtain nonperturbative solutions (RG trajectories) by truncating this theory space.
In the following we shall consider the Einstein-Hilbert approximation \cite{mr} whose truncated theory space is
2-dimensional. The corresponding ansatz for $\widehat{S}_k$ is parametrized by a $k$-dependent Newton constant
$G_k$ and cosmological constant $\bar{\lambda}_k$:
\be\label{9}  
\widehat{S}_k[g,\bar{g}] =
2\kappa Z_{Nk} \int d^d x\sqrt{g} \{ -R(g) +2\bar{\lambda}_k \}
+\kappa^2 Z_{Nk} \int d^d x\sqrt{g} \bar{g}^{\mu\nu} ( {\cal F}_\mu^{\alpha\beta}g_{\alpha\beta})
({\cal F}_\nu^{\rho\sigma}g_{\rho\sigma})
\ee
The first term on the RHS of (\ref{9}) is the familiar Einstein-Hilbert action in $d$-dimensions with the
$k$-dependent couplings. We wrote $G_k\equiv \bar{G}/{Z}_{Nk}$ for the running Newton constant, with a 
fixed constant $\bar{G}$, and introduced the convenient abbreviation $\kappa = (32 \bar{G})^{-1/2}$. The
second term of (\ref{9}) is the gauge fixing term for the background version of the harmonic coordinate 
condition \cite{adler,mr}, with gauge fixing parameter $\alpha=1$. The operator ${\cal F}_\mu^{\alpha\beta}$
is given by ${\cal F}_\mu^{\alpha\beta}= \delta_\mu^\beta\bar{g}^{\alpha\gamma}\bar{D}_{\gamma}
-\frac{1}{2}\bar{g}^{\alpha\beta}\bar{D}_\mu$ where the covariant derivative $\bar{D}_\mu$ is constructed from 
the background metric. Note that ${\cal F}_\mu^{\alpha\beta}\bar{g}_{\alpha\beta}=0$ so that the gauge fixing 
term drops from (\ref{9}) upon setting $g=\bar{g}$. The ghost Lagrangian pertaining to this ansatz \cite{mr}
is $\propto \bar{C}_\mu {{\cal M}^\mu}_\nu C^\nu$ with the kinetic operator ${\cal M}\propto S_{\rm gh}^{(2)}$
given by 
\be\label{10}
{\cal M}{[g,\bar{g}]^{\mu}}_\nu = \bar{g}^{\mu\rho}\bar{g}^{\sigma\lambda}\bar{D}_\lambda
(g_{\rho\nu} D_{\sigma}+g_{\sigma\nu}D_\rho)-\bar{g}^{\rho\sigma}\bar{g}^{\mu\nu}{\bar D}_\lambda
g_{\sigma\nu}D_\rho
\ee

Next we insert the truncation ansatz into the flow equation (\ref{10}). As a result, it boils down to a coupled 
system of ordinary differential equations for the two functions ${Z}_{Nk}$ and $\bar{\lambda}_k$.
They are obtained by performing a derivative expansion of the traces on the RHS of (\ref{8}) and then comparing 
the coefficients of the two invariants $\int\sqrt{g}$ and $\int\sqrt{g} R$ on both sides of the equation.
After having performed the second functional derivatives in $\widehat{S}^{(2)}_k$ we may set $\bar{h}=0$.
This leads to 
$\widehat{S}^{(2)}_k{[g,g]^{\mu\nu}}_{\rho\sigma} = 2\kappa^2 Z_{Nk}
[-{ K^{\mu\nu}}_{\rho\sigma}D^2+{U^{\mu\nu}}_{\rho\sigma}]$
where\\
\be\label{11}
{ K^{\mu\nu}}_{\rho\sigma}= \frac{1}{4}[\delta^\mu_\rho\delta^\nu_\sigma+\delta^\mu_\sigma\delta^\nu_\rho-g^{\mu\nu}g_{\rho\sigma}]
\ee
and 
\ba\label{12}
&&{U^{\mu\nu}}_{\rho\sigma} = 
\frac{1}{4}[\delta^\mu_\rho\delta^\nu_\sigma+\delta^\mu_\sigma\delta^\nu_\rho-g^{\mu\nu}g_{\rho\sigma}](R-2\bar{\lambda}_k)
+ \frac{1}{2}[ g^{\mu\nu}R_{\rho\sigma}+g_{\rho\sigma}R^{\mu\nu}]\nonumber\\[2mm]
&&-\frac{1}{4}[\delta^\mu_\rho {R^{\nu}}_\sigma+\delta^\mu_\sigma {R^{\nu}}_\rho+\delta^\nu_\rho {R^{\mu}}_\sigma
+\delta^\nu_\sigma {R^{\mu}}_\rho]
-\frac{1}{2}[{{{R^{\nu}}_\rho}^\mu}_\sigma + {{{R^\nu}_\sigma}^\mu}_\rho]
\ea 
It is sufficient to retain the terms proportional to $\int \sqrt{g}$ and $\int\sqrt{g} R$ from the
derivative expansion of the traces $\rm Tr \; exp (\cdots) $ appearing in (\ref{8}). They can be obtained 
straightforwardly by means of standard heat kernel techniques. Since the calculation is similar to
(but simpler than) the one in \cite{mr} we omit the details here. Also the subsequent 
$s$-integration can be performed easily in terms of Gamma functions. Equating the result to 
the LHS of the flow equation,
\be\label{13}
\partial_t \widehat{S}_k[g,g] =2\kappa^2 \int d^dx \sqrt{g} 
[ -R(g)\partial_t Z_{Nk} +2\partial_t (Z_{Nk}\bar{\lambda}_k)]
\ee 
we obtain two differential equations of the form $\partial_k Z_{Nk}=\cdots$ and $\partial_k(Z_{Nk}\bar{\lambda}_k)=\cdots$ .
It is convenient to rewrite them in terms of 
the dimensionless Newton constant $g(k)\equiv k^{d-2}G_k\equiv k^{d-2}Z_{Nk}^{-1}\bar{G}$
and the dimensionless cosmological constant $\lambda(k)\equiv k^{-2}\bar{\lambda}_k$. This leads to the following
system of equations:
\begin{subequations}
\ba\label{bb}
&&\partial_t g = \beta_g(g,\lambda) \equiv [d-2+\eta_N]g\\[2mm]
&&\partial_t \lambda = \beta_\lambda (g,\lambda)
\ea
\end{subequations}
The anomalous dimension $\eta_N \equiv -\partial_t {\rm ln} Z_{Nk}$ is given by
\be\label{15}
\eta_N=8(4\pi)^{1-\frac{d}{2}}\Big [ \frac{d(7-5d)}{24}\; (1-2\lambda)^{\frac{d}{2}-m-2}-\frac{d+6}{6}\Big ]g \; 
\frac{\Gamma(m+2-\frac{d}{2})}{\Gamma(m+1)}
\ee
and the beta-function of $\lambda$ reads
\be\label{16}
\beta_\lambda= -(2-\eta_N)\lambda +4(4\pi)^{1-\frac{d}{2}}\Big [\frac{d(d+1)}{4}\;(1-2\lambda)^{\frac{d}{2}-m-1}-d\Big ]g 
\frac{\Gamma(m+1-\frac{d}{2})}{\Gamma(m+1)}
\ee
The equations (15) with (\ref{15}) and (\ref{16}) are much simpler than their counterparts from the
average action. In particular they do not contain terms (proportional to $\eta_N$) stemming from the 
differentiation of the ${\cal Z}$-factors (see ref.\cite{mr}), and they allow for an explicit evaluation
of the threshold functions.
\section{universal quantities}
We have calculated the non-trivial fixed point $(g_\ast,\lambda_\ast) \not = 0$ in $d=4$, 
implied by the simultaneous vanishing of $\beta_g$ and $\beta_\lambda$ in (15), 
and the critical exponents $\theta'$, $\theta''$ \cite{oliver1,frank1} associated to the stability 
matrix for the non-trivial fixed point. The results are depicted in Tab.\ref{t1} for various 
values of the cut-off parameter $m$. The most important result is that the proper time
RG equation, too, predicts a non-Gaussian fixed point. It exists for any value of $m$.
The nonuniversal coordinates $\lambda_\ast$ and $g_\ast$ show a significant $m$-dependence, and it is
impressive to see how this $m$-dependence cancels out in the product $\lambda_\ast g_\ast$ which 
is universal in an exact calculation \cite{frank1}. For every value of $m$, the stability 
matrix $(-\partial_i \beta_j)$, $i,j \in \{g,\lambda \}$, has a pair of complex conjugate 
eigenvalues $\theta' \pm i \theta''$.  
\begin{table}
\caption{
Fixed point values and critical exponents for various values of the 
cutoff  parameter $m$. Note the plateau behavior of the universal 
quantities $\theta'$, $\theta''$, and  $\lambda_\ast g_\ast$ as the value of $m$ increases. \label{t1}} 
\begin{ruledtabular}
\begin{tabular}{cccccc}
$m$& $g_\ast$ & $\lambda_\ast$ &  $\lambda_\ast g_\ast$& $\theta'$& $\theta''$ \\
\hline
3/2& 0.763 & 0.192 & 0.147 & 2.000 & 1.658\\
2& 1.663 & 0.118 & 0.138 & 1.834 & 1.230\\
3& 1.890 & 0.066 & 0.125 & 1.769 &1.081 \\
4& 2.589 & 0.046 & 0.119 & 1.750 &1.001\\
5& 3.281 & 0.035 & 0.115 & 1.742 & 0.959\\
6& 3.970 & 0.028 & 0.113 & 1.737 &0.934\\
10& 6.718 & 0.016 & 0.108 & 1.729 & 0.886\\
40& 27.271 & 0.0038 & 0.103 & 1.722 & 0.840 \\
\end{tabular}
\end{ruledtabular}
\end{table}
Remarkably, those critical exponents and the  
product $\lambda_\ast g_\ast$ tend to constant values as $m$ increases; 
they form a ``plateau". The universal
quantities $\theta'$,$\theta''$, and $\lambda_\ast g_\ast$
show excellent stability properties 
for this class of regulators, and these values are
in complete agreement with the values found in the framework of 
the exact flow equation  
for the effective average action for gravity \cite{oliver1,frank1}. 
In fact, the differences between the proper time and the average action
RG equations are of the same order of magnitude as the residual scale dependence 
which is present in each of the two approaches and which is due to the truncated
theory space.
\section{conclusions}
We have presented a nonperturbative flow equation for gravity 
which employs a smooth cutoff function in the proper time
integration. The specific cutoff function employed in this paper   
is known to give rise to particularly accurate critical exponents 
in several scalar theories. The main result of our analysis is to have shown that
the non-Gaussian fixed point exists already in a simple improved 1-loop
calculation, an approach less sophisticated than that of the exact RG
equation.  

We have then shown that the universal quantities 
determined by the non-Gaussian fixed point are in very good agreement with the results from
the average action 
\cite{oliver1,frank1,oliver2}, and in particular, that they show a very weak dependence
on the regulator.  Our result can be taken as  
a further, conceptually different indication that the fixed point is rather ``robust" and should survive also
in a more general truncation.
\bibliography{pt}

\end{document}